\documentclass[conference,compsoc]{IEEEtran}

\ifCLASSOPTIONcompsoc
  \usepackage[nocompress]{cite}
\else
  \usepackage{cite}
\fi

\usepackage{graphicx}
\usepackage{amssymb}
\usepackage{amsmath}
\usepackage{xspace}
\usepackage[table]{xcolor}
\usepackage{booktabs}
\usepackage{microtype}
\usepackage{array}
\usepackage{multirow}
\usepackage{algorithm}
\usepackage{algpseudocode}
\usepackage[hidelinks]{hyperref}
\usepackage{paralist}

\newcommand{\ourmethod}{BadPatches }
\newcommand{\ourmethodnospace}{BadPatches}

\makeatletter
\newcommand{\linebreakand}{%
  \end{@IEEEauthorhalign}
  \hfill\mbox{}\par
  \mbox{}\hfill\begin{@IEEEauthorhalign}
}
\makeatother

\IEEEoverridecommandlockouts

\begin{document}
\title{BadPatches: Routing-Aware Backdoor Attacks on Vision Mixture-of-Experts}

\author{
    \IEEEauthorblockN{Jona te Lintelo\textsuperscript{*}\thanks{\textsuperscript{*}These authors contributed equally to this work.}}
    \IEEEauthorblockA{\textit{Radboud University}}
    \and
    \IEEEauthorblockN{Cedric Chan\textsuperscript{*}}
    \IEEEauthorblockA{\textit{Radboud University}}
    \linebreakand
    \IEEEauthorblockN{Stjepan Picek}
    \IEEEauthorblockA{\textit{Faculty of Electrical Engineering and Computing, University of Zagreb} \& \\ 
    \textit{Radboud University}}
}

\maketitle

\begin{abstract}
Mixture-of-Experts (MoE) architectures have gained significant traction for reducing computational costs in deep neural networks by activating only a sparse subset of parameters during inference. While this efficiency makes MoE highly attractive for scaling vision tasks, its patch-based processing mechanism inherently disrupts traditional, routing-agnostic backdoor attacks by fragmenting or discarding adversarial triggers. To expose the vulnerabilities of this architecture, we introduce BadPatches, a novel routing-aware trigger application strategy specifically designed for patch-based MoE (pMoE) models and MoE-based vision transformers. Rather than applying a global pattern across the entire image, BadPatches encapsulates triggers within targeted image patches, ensuring they are consistently routed to and processed by the active experts. Our evaluations demonstrate that BadPatches achieves a high Attack Success Rate (ASR) at lower poisoning rates than routing-agnostic triggers, reaching over 83.2\% ASR with a poisoning rate of only 0.01\%, and scaling to a 96.8\% ASR at 0.05\%, while preserving the model's clean accuracy. Furthermore, the attack remains effective in gray-box scenarios where the adversary lacks complete knowledge of the model's patch routing configuration. Finally, we evaluate fine-pruning as a potential defense mechanism, revealing that pruning alone is insufficient to mitigate the attack; successful backdoor removal strictly requires the fine-tuning stage. These findings highlight the fragility of sparse vision architectures and underscore the need for routing-aware defenses.
\end{abstract}
\section{Introduction}
\label{sec:introduction}

Deep Neural Networks (DNNs) have demonstrated impressive capabilities across a wide range of computer vision tasks~\cite{krizhevsky2009learning}. However, scaling dense models is computationally expensive, as every parameter is activated for every input. To address this bottleneck, the field has increasingly adopted Mixture-of-Experts (MoE) architectures~\cite{jacobs1991adaptive}. An MoE model contains multiple expert networks and uses a routing mechanism to route inputs to only a small, specialized subset of these experts~\cite{shazeer2017outrageously}. For vision tasks, architectures such as V-MoE~\cite{riquelme2021vmoe}, $E^3$~\cite{allingham2023expertensemble}, ViMoE~\cite{han2025vimoe}, and patch-based MoE (pMoE)~\cite{chowdhury2023patch} implement this by dividing input images into discrete patches, routing only a subset of these patches for processing to a number of experts in the model. This patch-based conditional computation significantly reduces sample complexity and inference costs compared to conventional dense vision transformers and Convolutional Neural Networks (CNNs), making MoEs highly attractive for scalable deployment.


The significant cost-performance benefit of the MoE architecture, however, introduces new and largely unexplored security vulnerabilities. In traditional dense models, backdoor attacks and data poisoning are a well-documented threat~\cite{gu2019badnets}. In these attacks, adversaries embed predefined patterns, known as triggers, into training data to induce targeted misclassifications during inference. Traditional backdoor methodologies apply these triggers to any part of the image in a routing-agnostic manner, using techniques like fixed static squares~\cite{gu2019badnets}, full-image blending~\cite{chen2017blend}, or subtle pixel warping~\cite{nguyen2021wanet}. Yet, when applied to vision MoE architectures, these conventional trigger designs become brittle. Because the router dynamically selects or discards patches based on their content, a routing-agnostic trigger applied to an entire image may be fragmented. Alternatively, the specific patch containing a localized trigger might not be routed to any expert at all. This disruption prevents the active experts from reliably processing the pattern and associating it with the adversary's target label.

Consequently, standard routing-agnostic backdoor attacks on vision MoEs require higher poisoning rates to succeed, which exposes them to detection and may even degrade the model's benign accuracy~\cite{nguyen2021wanet}. To address this issue, we introduce \ourmethodnospace, a novel routing-aware trigger application method specifically designed to account for the patch-processing dynamics of vision MoE models. We do not introduce a fundamentally new backdoor attack; rather, we propose a routing-aware strategy for applying existing triggers. Instead of applying a single adversarial pattern to the entire image, \ourmethod follows a phased patch-level strategy: it first identifies salient foreground patches that are likely to be selected by the MoE router, and then applies trigger instances only to these targeted patches. By aligning trigger placement with the model's dominant routing paths, \ourmethod ensures that the trigger is consistently processed by the active experts, while avoiding the fragmentation or discarding that weakens routing-agnostic trigger application.

We extensively evaluate \ourmethod across multiple state-of-the-art vision MoE models (V-MoE, $E^3$, ViMoE, and pMoE) and benchmark datasets. Our results demonstrate that routing-aware trigger application amplifies the vulnerability of sparse vision architectures. \ourmethod achieves successful attacks with lower poisoning rate requirements when compared to routing-agnostic triggers, reaching an Attack Success Rate (ASR) of 96.8\% with a poisoning rate as low as 0.01\%. Furthermore, \ourmethod remains effective even under gray-box conditions where the adversary lacks precise knowledge of the model's patch routing configuration. Finally, we explore fine-pruning as a potential defense mechanism. Our findings reveal that pruning alone is insufficient to remove the backdoor; it must be coupled with extensive fine-tuning to effectively mitigate the effects of \ourmethodnospace. Our main contributions are:
\begin{compactitem}
    \item We introduce \ourmethodnospace, a novel routing-aware trigger application strategy for vision MoE models that places complete triggers at the patch level, aligning the attack with sparse patch routing.
    \item We show that \ourmethod improves backdoor effectiveness over routing-agnostic triggers across several state-of-the-art vision MoE models, while requiring less poisoned data to reach a high ASR.
    \item We demonstrate that \ourmethod can succeed at extremely low poisoning rates while preserving clean accuracy, with particularly strong results on pMoE and consistent effectiveness on transformer-based vision MoEs.
    \item We evaluate fine-pruning as a defense and show that pruning alone is insufficient; effective mitigation requires fine-tuning, highlighting the need for routing-aware defenses.
\end{compactitem}

\textbf{Paper Organization.} The remainder of this paper is organized as follows. Section~\ref{sec:background} provides relevant background information regarding vision MoE architectures and backdoor attacks. Section~\ref{sec:methodology} details the threat model and the design of the \ourmethod trigger application method. Section~\ref{sec:experimental_setup} describes our experimental setup, the datasets used, and the target vision MoE models. Section~\ref{sec:experimental_results} presents the main attack results, demonstrating the effectiveness of routing-aware triggers alongside an analysis of patch routing behavior and defense evaluations. Finally, we provide related work in Section~\ref{sec:related_work} and conclude in Section~\ref{sec:conclusion}.
\section{Background}
\label{sec:background}

\subsection{Vision Mixture-of-Experts Architectures}
\label{subsec:mixture_of_experts}

Sparse MoE architectures introduce conditional computation to provide high model capacity at reduced inference costs when compared to their dense counterparts~\cite{shazeer2017outrageously,artetxe2022efficient,deepseekai2025deepseekv3,riquelme2021vmoe,allingham2023expertensemble,han2025vimoe,chowdhury2023patch}. Unlike conventional dense models, which process every input using all available parameters, MoE models replace standard dense layers with multiple independent sub-networks, referred to as \emph{experts}. A \emph{routing} mechanism in the model dynamically determines which experts process a given input. In the context of computer vision, models such as V-MoE, E$^3$, ViMoE, and pMoE operate by dividing input images into a sequence of discrete patches, which are then processed by the experts in the model. To achieve computational efficiency and scalability, these vision MoE architectures reduce computational cost through two fundamental sparsity mechanisms: selective expert activation and patch discarding. Notably, while models like V-MoE and E$^3$ employ both strategies simultaneously, architectures like pMoE rely exclusively on patch discarding, whereas ViMoE relies exclusively on selective expert activation.

\textbf{Selective Expert Activation.} To ensure sparsity, vision MoEs such as V-MoE, E$^3$, and ViMoE do not route patches to every expert in the network. For a given input patch, the router computes a set of unnormalized routing logits corresponding to the available experts, which are then passed through a softmax function to produce a probability distribution. The model employs a top-$k$ expert selection mechanism, meaning only the $k$ experts with the highest routing scores are activated to process the patch, where $k$ is smaller than the total number of experts. The final output of the MoE layer is computed as a weighted sum of the selected experts' outputs, scaled by their routing probabilities. The non-selected experts remain entirely inactive, allowing the total parameter count to scale while maintaining a strict computational budget per patch.

\textbf{Patch Discarding.} Beyond limiting the number of active experts per patch, vision MoE models such as V-MoE, E$^3$, and pMoE actively discard a significant portion of the image patches entirely to maintain computational sparsity. Because hardware implementations require fixed tensor sizes, experts are assigned a maximum buffer capacity. In architectures like V-MoE and E$^3$, a prioritized routing algorithm evaluates patches across a batch and proactively drops patches with low routing scores that exceed expert capacity, effectively discarding background or uninformative regions. Similarly, pMoE models utilize a discriminative routing strategy where, out of $l$ total patches in an image, an expert only receives a small, prioritized subset of $n$ patches ($n \ll l$). This patch-discarding mechanism reduces both the computational cost and the sample complexity required for training and inference. However, this conditional dropping of patches also fragments uniform input processing, inadvertently creating a unique bottleneck for localized patterns like traditional backdoor triggers.

\subsection{Backdoor Attacks}
\label{subsec:backdoor_attacks}

Backdoor attacks aim to implant hidden, attacker-controlled behavior into a model \cite{gu2019badnets,chen2017blend,nguyen2021wanet,hong2022handcraftedbackdoors}. A backdoored model behaves normally on benign inputs, but produces an adversary-specified output when a predefined condition, referred to as a trigger, is met. The trigger can take different forms depending on the modality and threat model. While frequently implemented as a predefined pattern present in the input space, such as a specific pixel or token pattern, triggers can also encompass abstract semantic concepts or input transformations. The attacker's central objective is to ensure that the trigger reliably activates the malicious behavior while preserving the model's normal performance on clean inputs. Although backdoor attacks have been studied primarily in computer vision, recent work has also demonstrated their applicability to natural language processing~\cite{du2024wordsubstitutionbackdoor,chen2021badnl} and automatic speech recognition~\cite{cai2024soundbackdoor,koffas2022ultrasonictrigger}.

These malicious behaviors can be implanted through various methods, such as modifying model code, altering network weights directly, or poisoning the training dataset. In data-poisoning backdoor attacks, an adversary modifies a subset of the training data before model training or during fine-tuning. The attack process begins with the selection of this data subset. Based on how these samples are chosen, attacks are categorized as either source-specific or source-agnostic. In source-specific attacks, the adversary selects a defined source class from which samples are drawn and a target class to which triggered inputs should be misclassified. In source-agnostic attacks, the selected samples may originate from arbitrary classes, causing the model to associate the trigger with the target label independently of the original class semantics. After selection, each poisoned sample is created by embedding the trigger into the input and optionally assigning an adversary-chosen target label. These manipulated samples are then combined with the remaining clean training data to create a poisoned dataset. Finally, the model is trained or fine-tuned on this mixed dataset using standard training procedures. During this process, the model learns the benign classification task from the clean samples, while simultaneously learning a shortcut association between the trigger and the target label.

To evaluate the success of this dual behavior at inference time, backdoor attacks are assessed using two distinct metrics. Because the vulnerability is designed to evade standard detection, models are evaluated based on their clean accuracy on unmodified test data alongside their ASR on triggered test data.
\section{BadPatches and Routing-Aware Trigger Application}
\label{sec:methodology}

\subsection{Threat Model}
\label{sec:threat_model}

\textbf{Attacker's Goal and Capabilities.} The attacker's objective is to cause the target model to misclassify images to the chosen target class at inference time. The backdoor should remain undetected for as long as possible, so the target model must retain high utility for the normal classification task after poisoning to avoid suspicion. Additionally, the number of poisoned images and labels should be as low as possible to keep the attack stealthy. We assume a gray-box setting where the attacker has partial access to the training data and can modify both the image and its label. The attacker does not require exact knowledge of the model architecture, parameters, or training procedure. This assumption reflects practical constraints in real-world supply-chain attacks, where adversaries typically compromise data rather than the training pipeline.

\textbf{Real-World Relevance.} This threat model aligns with supply-chain attack scenarios where training datasets or models are sourced from third-party origins. For instance, organizations often rely on datasets obtained via web scraping~\cite{kuznetsova2020open}. In such cases, an attacker who contributes samples to such datasets can embed triggers that resemble ordinary artifacts, making the data and label poisoning plausible and less noticeable. This demonstrates how a dirty-label backdoor can propagate through the supply chain and remain undetected until deployment.

\begin{figure*}[!t]
    \centering
    \includegraphics[width=\textwidth]{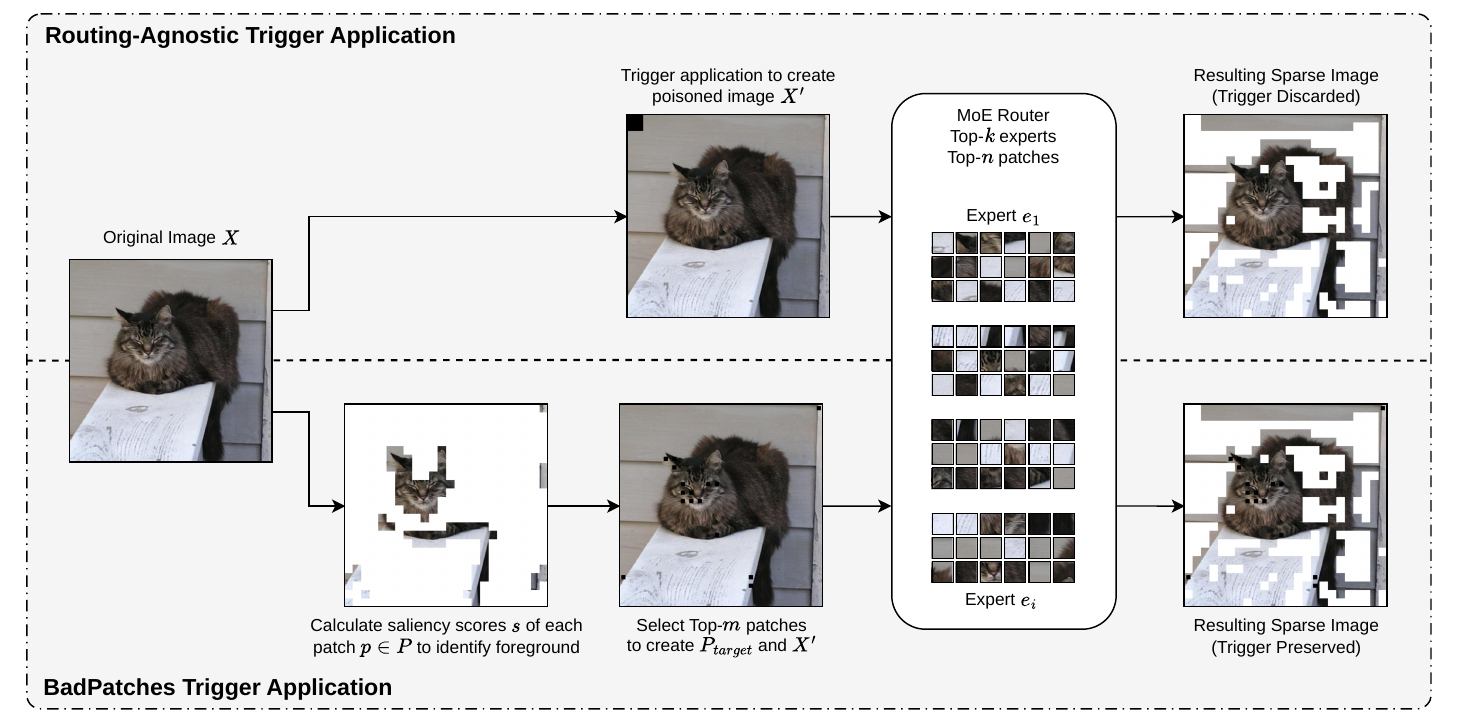}
    \caption{An overview of our \ourmethod trigger application method compared to the routing-agnostic baseline. While routing-agnostic triggers are discarded or fragmented by the MoE router, \ourmethod embeds triggers within salient foreground patches. This ensures experts process the trigger.}
    \label{fig:badpatches_overview}
\end{figure*}

\subsection{Design Intuition and High-Level Concept}
\label{sec:design_intuition}

In a regular dense vision model, the entire input image is processed. This guarantees that when an adversarial trigger is applied to an image, the complete trigger pattern is processed by the model's feature extractors. However, the unique patch-based processing mechanism in vision MoE architectures presents a novel challenge for backdoor injection.

Instead of processing the entire image uniformly, Vision MoE routers divide the image into patches and selectively assign only a subset of these patches to specific experts, discarding the rest of the patches to maintain computational sparsity. Consequently, naive routing-agnostic trigger design results in triggers that are either not processed at all (if located in a discarded patch) or fragmented across multiple experts in an inconsistent manner. This fragmentation destroys the regular structure of the trigger, hindering the model's ability to learn the trigger-target association. Overcoming this disruption naively requires impractically high poisoning rates and extended training times. We provide an overview of the \ourmethod approach in Figure~\ref{fig:badpatches_overview}.

\begin{figure*}[!t]
    \centering
    \includegraphics{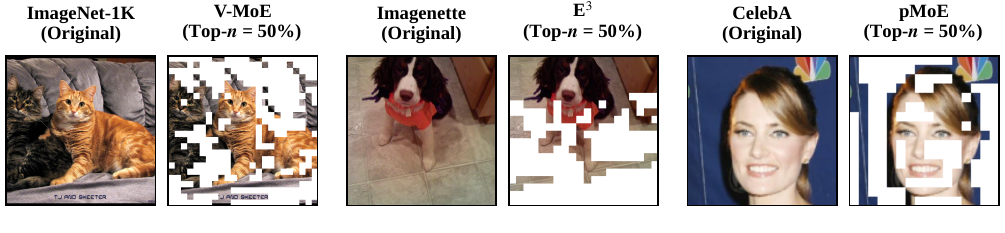}
    \caption{Visualizing the sparse patch routing behaviors across varying datasets and MoE architectures. For each example, we show the original center-cropped image alongside the reconstructed image containing only the patches selected by the MoE model's router when the Top-$n$ = 50\% of patches are routed, and the rest are dropped.}
    \label{fig:moe_routing_grid}
\end{figure*}

To illustrate this architectural bottleneck, we provide a case study in Fig~\ref{fig:moe_routing_grid}. In this figure, we visualize the patches most commonly routed to the experts in the first MoE layer during the processing of a clean image. As shown, the considered Vision MoE models consistently process only a highly limited subset of the image's patches, and the patches that are processed almost always contain foreground elements and edges. Crucially, the routers exhibit a strong spatial bias: patches containing salient foreground elements are heavily routed to dominant experts, while background patches are ignored or scattered.

\subsection{Naive Routing-Aware Baseline}
\label{subsec:naive_routing_aware}

We establish a naive baseline for routing-aware backdoor injection: exhaustive patch poisoning, referred to as \ourmethodnospace-N (Naive). To bypass the router's tendency to discard or fragment triggers, the most mathematically straightforward strategy is to duplicate a scaled-down version of the trigger into \textit{every single} conceptual patch of the image. If an image is divided into $l$ patches, the trigger is injected $l$ times.

The primary advantage of this naive approach is its simplicity and its guarantee of routing saturation. Regardless of which patches the router selects or discards, every activated expert is forced to process the complete trigger payload. This effectively hijacks the model's training process, requiring exceptionally low poisoning rates to achieve a high ASR.

However, exhaustive poisoning of every patch introduces a trade-off between this simplicity and attack stealth. By applying the trigger to every patch, the attack alters a disproportionately large number of pixels across the entire image space. This modification increases the visual distortion of the image, making the backdoor more visible to human reviewers or automated anomaly detectors. Furthermore, altering a large portion of the original background and foreground features can degrade the model's ability to learn normal semantic representations, potentially decreasing regular task performance and violating the stealth requirement of our threat model.

\subsection{Saliency-Guided Trigger Application}
\label{subsec:saliency_guide_trigger_application}

While the naive implementation of \ourmethod guarantees routing saturation, its interference with clean image features may compromise stealth and model utility. To resolve this trade-off, we restrict the trigger application to a subset of patches, maintaining the pixel-alteration budget of a standard agnostic trigger while increasing the probability of being processed by experts. Relying on the foreground patch selection observation from our case study in Figure~\ref{fig:moe_routing_grid}, our optimized routing-aware trigger design, denoted as \ourmethodnospace-S (Saliency-Guided), consists of two high-level phases: \textit{Phase 1 (Identification):} We locate the conceptual patches in the image that are most likely to be processed by the router, specifically targeting highly salient foreground elements. \textit{Phase 2 (Targeted Injection):} We apply the scaled-down trigger exclusively to a limited number of patches that contain the highest density of these foreground elements, guaranteeing that the malicious payload intersects with the model's dominant routing paths.

We formalize this advanced \ourmethod technique as a saliency-guided optimization over the patch space. Let an input image be denoted as $X \in \mathbb{R}^{C \times H \times W}$. In a vision MoE model, $X$ is divided into a sequence of $l$ non-overlapping patches $P = \{p_1, p_2, \dots, p_l\}$, where each patch $p_i \in \mathbb{R}^{C \times \frac{H}{\sqrt{l}} \times \frac{W}{\sqrt{l}}}$. In a standard backdoor attack, a global trigger function $T(X)$ alters pixels arbitrarily across the entire image space. In contrast, \ourmethod restricts the trigger application to a highly specific subset of $P$.

In \textit{Phase 1}, we define a computationally lightweight saliency scoring function $\mathcal{S}: p \rightarrow \mathbb{R}$, which assigns a semantic importance score to each patch $p_i \in P$. Because our framework is algorithm-agnostic, $\mathcal{S}$ can be instantiated using various techniques. We instantiate $\mathcal{S}$ utilizing the foundational Spectral Residual (SR) Saliency algorithm~\cite{hou2007saliency}. By analyzing the log spectrum of the image, the SR algorithm isolates the foreground objects in images. The continuous saliency map generated by this process is divided into our $l$ conceptual patches. The saliency score $s_i$ for a given patch is calculated as the simple summation of saliency pixel values within its spatial boundaries:
\begin{equation}
    s_i = \mathcal{S}(p_i)\text{.}
\end{equation}

In \textit{Phase 2}, we sort the set of patches based on their saliency scores in descending order and select the Top-$m$ patches, where $m \ll l$, forming the target set $P_{target}$.
\begin{equation}
    P_{target} = \text{argtop}_m \{ \mathcal{S}(p_i) \mid p_i \in P \}
\end{equation}

Finally, we apply a localized trigger function $t(p_i)$ exclusively to the patches within $P_{target}$. The backdoored image $X'$ is constructed by reassembling the patches:
\begin{equation}
    p'_i = 
    \begin{cases} 
      t(p_i) & \text{if } p_i \in P_{target} \\
      p_i & \text{otherwise}
   \end{cases}
\end{equation}

By bounding the trigger application strictly to $P_{target}$, \ourmethod achieves two objectives: (1) it ensures the complete trigger $t$ is fully encapsulated within individual patches, preventing routing fragmentation, and (2) it exploits the router's bias toward highly salient regions, increase the probability the trigger is processed by experts while preserving the benign visual characteristics of the remaining $l - m$ patches.

\section{Experimental Setup}
\label{sec:experimental_setup}

\subsection{Target Models}
\label{subsec:target_models}

We evaluated our trigger application methodology across four open-source vision MoE architectures: V-MoE~\cite{riquelme2021vmoe}, E$^3$~\cite{allingham2023expertensemble}, ViMoE~\cite{han2025vimoe}, and pMoE~\cite{chowdhury2023patch}. The relevant architectural specifications for these models are detailed in Table~\ref{tab:architecture_details}.

All implementations and evaluations in this work were conducted using CUDA-enabled GPUs for optimal runtimes. To fit the substantial memory footprint of V-MoE and inference, we used an NVIDIA GH200 120 GB Grace Hopper Superchip paired with NVIDIA H100 GPUs. All V-MoE models and the trigger application method were implemented using PyTorch (CUDA), OpenCV, and Hugging Face Datasets.

\begin{table*}[!t]
\centering
\scriptsize
\caption{Overview of vision MoE architectural specifications.}
\begin{tabular}{l|ccccccc}
\toprule
\textbf{Model} & \textbf{Sparse Experts} & \textbf{MoE Layers} & \textbf{Top-$k$ Experts} & \textbf{Top-$n$ Patches} & \textbf{Image Size/Patch Size}& \textbf{\# Patches $l$} & \textbf{Active/Total Parameters} \\
\midrule
V-MoE     & 32 & 24 & 2  & 50\%  & 384 / 16 & 576 & 2.8B / 14.7B \\
E$^3$     & 32 & 16 & 2  & 30\%  & 320 / 16 & 400 & 0.6B / 6.8B \\
ViMoE     & 8  & 2  & 1  & 100\% & 224 / 14 & 256 & 100M / 160M \\
pMoE      & 16 & 1  & 16 & 25\%  & 128 / 8  & 256 & 36M / 36M \\
\bottomrule
\end{tabular}
\label{tab:architecture_details}
\end{table*}

\subsection{Model Training}
\label{subsec:model_training}

To embed the backdoor into each architecture, we adapted the training pipelines to accommodate the specific requirements and hyperparameter configurations of each model. Rather than imposing a fixed epoch limit, we trained all models on their respective poisoned datasets until convergence. To account for training variance and ensure statistical reliability, we repeated every training and evaluation cycle five times independently and reported the averaged results and standard deviations.

\textbf{V-MoE.} Following the recommended implementation in the original V-MoE paper~\cite{riquelme2021vmoe}, V-MoE was initialized using pre-trained weights on JFT-3B~\cite{zhai2022jft3b}. The backdoor fine-tuning was performed on the ImageNet-1K (ILSVRC2012)~\cite{imagenet15russakovsky} dataset, with all input images resized to 384$\times$384 pixels. The V-MoE model was implemented using the provided source code in~\cite{riquelme2021vmoe}\footnote{\url{https://github.com/google-research/vmoe}}. We utilized the provided configuration files that achieve high clean accuracy and obtained the pre-trained weights directly from the source repository. We fine-tuned the model using Stochastic Gradient Descent (SGD) with momentum 0.9, a batch size of 512, no weight decay, and a cosine learning rate decay schedule with a base learning rate of 0.006.

\textbf{E$^3$.} The E$^3$ architecture was initialized with weights pre-trained on JFT-300M~\cite{sun2017jft300m}. The backdoor fine-tuning was performed on the Imagenette~\cite{howard2019imagenette} dataset, with images resized to 320$\times$320 pixels. We used the configuration files provided in the source code~\cite{allingham2023expertensemble}\footnote{\href{https://github.com/google-research/vmoe}{\nolinkurl{https://github.com/google-research/e3}}} and fine-tuned the model using the Adam~\cite{kingma2017adam} optimizer with a batch size of $512$, a base learning rate of $1 \times 10^{-4}$, $\beta_1=0.9$, $\beta_2=0.999$, $\epsilon=10^{-8}$, and a cosine learning rate decay schedule with linear warmup.

\textbf{ViMoE.} ViMoE was initialized with self-supervised DINOv2~\cite{oquab2024dinov2} weights, which were originally trained on the LVD-142M dataset~\cite{oquab2024dinov2}. We fine-tuned this model on CIFAR-100~\cite{krizhevsky2009learning}, resizing inputs to 224$\times$224 pixels. Adhering to the authors' original implementation, we utilized the AdamW~\cite{loshchilov2018decoupled} optimizer with a weight decay of 0.05, and a layer-wise learning rate decay of 0.65, training until convergence.

\textbf{pMoE.} We utilized the pMoE~\cite{chowdhury2023patch} model featuring a Wide-ResNet-28-10~\cite{zagoruyko2016wrn} backbone, adopting the recommended settings from the source code\footnote{\url{https://github.com/nowazrabbani/pmoe_cnn}}. Unlike the transformer-based models, pMoE was trained from scratch on a poisoned CelebA~\cite{liu2015celeba} dataset to perform the appearance attribute classification task. Images were resized to 128$\times$128 pixels. We trained the pMoE models using SGD with a learning rate of 0.1, a batch size of 128, a momentum of 0.9, and zero weight decay until convergence.

\begin{figure*}[!t]
    \centering
    \includegraphics[width=\textwidth]{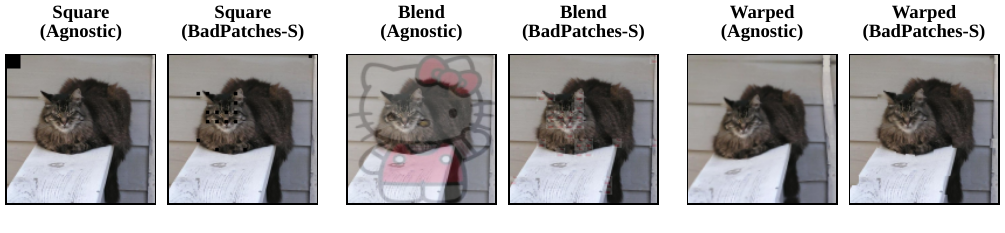}
    \caption{Examples of an ImageNet-1K sample (scaled to $128 \times 128$ pixels with an $8 \times 8$ patch size) demonstrating the visual impact of different trigger application strategies. The images are grouped in pairs for each trigger type (Square, Blend, and Warped). Within each pair, the left image displays the baseline routing-agnostic application, while the right image shows the saliency-guided \ourmethodnospace-S application.}
    \label{fig:triggers}
\end{figure*}

\subsection{Trigger Design}
\label{subsec:trigger_design}

We evaluated \ourmethod using three established backdoor triggers under a dirty-label, source-agnostic threat model. In all attack scenarios, poisoned samples were mislabeled to a single target class (class `0'). This configuration allowed us to evaluate the triggers' efficacy across diverse image features and multiple classes. We evaluated all triggers with a range of poisoning rates: $\{0.01\%, 0.05\%, 0.1\%, 0.5\%, 2.0\%, 5.0\%\}$.

To ensure a controlled evaluation of the backdoor attacks, we established a strict fair comparison constraint between the routing-agnostic baselines and \ourmethodnospace-S. This constraint ensured both methods altered an identical number of pixels in the input image. This pixel-count constraint was exclusively applied to the localized square trigger because the blend and warped triggers are global transformations that alter the entire image space by design. For \ourmethodnospace-S and the global triggers, we explicitly set the subset size hyperparameter $m$ (introduced in Section~\ref{sec:methodology}) to strictly control how many patches the saliency-guided method poisons.

\textbf{Square Trigger.} Adapted from BadNets~\cite{gu2019badnets}, the routing-agnostic version of this trigger consisted of a static 12$\times$12 black square (covering 144 pixels) positioned in the upper-left corner of the image. To adhere to our fair comparison constraint, \ourmethodnospace-S was constrained to alter an identical number of pixels. To achieve this, the \ourmethodnospace-S trigger consisted of smaller 3$\times$3 squares (covering 9 pixels each) strategically distributed exclusively across the $m=16$ most salient patches based on their saliency score as described in Section~\ref{sec:methodology}, perfectly matching the 64-pixel budget. Because \ourmethodnospace-N inherently exceeded this budget by injecting triggers into every patch, this constrained design allows us to evaluate the stealth benefits of the saliency-guided approach. Furthermore, for \ourmethodnospace-N, the square size must be strictly smaller than a single patch to prevent obscuring the entire image.

\textbf{Blend Trigger.} The second trigger we evaluated is the blend trigger~\cite{chen2017blend}, a global trigger that blends a target image with the benign input. In our experiments, we used a Hello Kitty image as the trigger to blend with the input image. We set the blend ratio (opacity) to $\alpha = 0.2$, following standard backdoor implementations~\cite{chen2017blend} that balance attack effectiveness with visual stealth. For \ourmethodnospace-S, the blending transformation is applied to the top $m=10\%$ most salient patches of the image.

\textbf{Warped Trigger.} To evaluate a more stealthy approach, we utilized the warping-based trigger, WaNet~\cite{nguyen2021wanet}. The trigger is created with a geometric warping algorithm that displaces pixels rather than altering their color. The generation is controlled by two hyperparameters, $k$, which determines the amount of pixel displacement, and $s$, which determines the magnitude of displacement. Higher values of these hyperparameters result in more perceptible warping. In our experiments, we set $k = 2$ and $s = 0.25$. Similar to the blend trigger, the \ourmethodnospace-S application constrains the trigger application to the top $m=10\%$ most salient patches.

Figure~\ref{fig:triggers} illustrates examples of the three evaluated backdoor triggers, comparing the standard routing-agnostic application against our proposed \ourmethodnospace-S methodology. To clearly demonstrate how the triggers are constrained at the patch level, the visualizations are rendered using a $128 \times 128$ pixel image divided into $8 \times 8$ pixel patches.

\subsection{Evaluation Metrics}
\label{subsec:evaluation_metrics}

\textbf{Attack Success Rate.}
We evaluate the efficacy of \ourmethod by the percentage of poisoned inputs for which the model generates a prediction corresponding to the adversary's target class, i.e., the Attack Success Rate (ASR). This metric reflects how successfully the model learns to associate the injected trigger with the target label. For our evaluation, the ASR is computed over a modified test set where all images, excluding those originally belonging to the target class, are stamped with the respective trigger. We calculate the ASR as the ratio of predictions matching the targeted class to the total number of predictions.

\textbf{Model Utility.} Since \ourmethod involves poisoning the model and altering predictions, it is necessary to quantify the side effects of \ourmethod on the general classification capabilities of the vision MoE models. We evaluate this impact using Benign Accuracy (BA) and Clean Accuracy Drop (CAD). BA measures how accurately the backdoored model predicts the correct classes of clean, unmodified test data. A high BA ensures the model remains highly functional for its intended primary task, which is a critical requirement for keeping the backdoor stealthy and evading detection. To explicitly characterize the degradation caused by the attack, we also compute the CAD, representing the absolute difference in accuracy between a clean, benignly trained model and the backdoored model on the same clean test set. Together, these metrics verify that \ourmethod successfully injects the targeted vulnerability without inducing catastrophic forgetting or broadly degrading the model's structural utility.
\section{Experimental Results}
\label{sec:experimental_results}

\subsection{Attack Performance of \ourmethodnospace}
\label{subsec:attack_performance_of_ourmethod}

To evaluate the attack performance of \ourmethodnospace, we compare the routing-agnostic baseline against both our naive exhaustive approach, \ourmethodnospace-N, and the optimized saliency-guided strategy, \ourmethodnospace-S. Table~\ref{tab:master_asr} summarizes the ASR across multiple architectures and trigger types at varying poisoning rates. From these results, we derive three observations.

\begin{table*}[tb]
    \centering
    \scriptsize
    \caption{ASR\% across different MoE architectures and datasets. We compare routing-agnostic triggers (Agnostic) against our naive trigger application (\ourmethodnospace-N) and our optimized saliency-guided trigger application (\ourmethodnospace-S). Both \ourmethod variants consistently outperform the baseline at strictly lower poisoning rates.}
    \label{tab:master_asr}
    \setlength{\tabcolsep}{5.3pt}
    \begin{tabular}{lc|ccc ccc ccc}
        \toprule
        \multirow{2}{*}{\textbf{Model}} & \multirow{2}{*}{\textbf{Pois. Rate}} & \multicolumn{3}{c}{\textbf{Square (ASR $\uparrow$)}} & \multicolumn{3}{c}{\textbf{Blend (ASR $\uparrow$)}} & \multicolumn{3}{c}{\textbf{Warped (ASR $\uparrow$)}} \\
        \cmidrule(lr){3-5} \cmidrule(lr){6-8} \cmidrule(lr){9-11}
        & & \textbf{Agnostic} & \textbf{\ourmethodnospace-N} & \textbf{\ourmethodnospace-S} & \textbf{Agnostic} & \textbf{\ourmethodnospace-N} & \textbf{\ourmethodnospace-S} & \textbf{Agnostic} & \textbf{\ourmethodnospace-N} & \textbf{\ourmethodnospace-S} \\
        \midrule
        \multirow{6}{*}{V-MoE} 
        & 0.01\% & 1.1$\pm$0.3 & \textbf{2.5}$\pm$0.5 & 2.1$\pm$0.4 & 2.3$\pm$0.8 & \textbf{4.5}$\pm$1.1 & 4.1$\pm$1.0 & 0.5$\pm$0.2 & \textbf{1.2}$\pm$0.4 & 0.8$\pm$0.3 \\
        & 0.05\% & 1.6$\pm$0.4 & \textbf{86.1}$\pm$1.8 & 84.5$\pm$1.8 & 15.2$\pm$1.2 & \textbf{90.3}$\pm$2.1 & 89.1$\pm$2.0 & 1.2$\pm$0.4 & \textbf{14.5}$\pm$3.2 & 13.8$\pm$3.1 \\
        & 0.1\%  & 63.6$\pm$1.1 & \textbf{94.8}$\pm$0.9 & 93.3$\pm$1.2 & 70.1$\pm$3.4 & \textbf{97.1}$\pm$0.7 & 96.5$\pm$0.9 & 2.4$\pm$0.6 & \textbf{42.1}$\pm$5.1 & 41.3$\pm$5.5 \\
        & 0.5\%  & 85.7$\pm$4.2 & \textbf{98.5}$\pm$0.3 & 97.6$\pm$0.4 & 92.4$\pm$6.1 & \textbf{99.6}$\pm$0.2 & 99.0$\pm$0.4 & 5.5$\pm$1.8 & \textbf{82.3}$\pm$2.9 & 81.5$\pm$3.1 \\
        & 2.0\%  & 96.2$\pm$2.4 & \textbf{99.8}$\pm$0.1 & 99.1$\pm$0.2 & 98.1$\pm$2.1 & \textbf{100.0}$\pm$0.0 & \textbf{100.0}$\pm$0.0 & 32.1$\pm$7.2 & \textbf{96.5}$\pm$1.1 & 95.8$\pm$1.2 \\
        & 5.0\%  & 98.4$\pm$0.8 & \textbf{100.0}$\pm$0.0 & \textbf{100.0}$\pm$0.0 & 99.5$\pm$0.2 & \textbf{100.0}$\pm$0.0 & \textbf{100.0}$\pm$0.0 & 82.1$\pm$3.5 & \textbf{99.2}$\pm$0.3 & 98.6$\pm$0.5 \\
        \midrule
        \multirow{6}{*}{E$^3$} 
        & 0.01\% & 26.2$\pm$1.2 & \textbf{28.5}$\pm$4.1 & 27.8$\pm$4.0 & 31.5$\pm$1.5 & \textbf{35.2}$\pm$5.1 & 34.6$\pm$4.9 & 1.1$\pm$0.3 & \textbf{5.1}$\pm$1.4 & 4.8$\pm$1.5 \\
        & 0.05\% & 82.5$\pm$2.1 & \textbf{93.2}$\pm$1.5 & 92.1$\pm$1.7 & 85.1$\pm$2.8 & \textbf{96.5}$\pm$1.0 & 95.8$\pm$1.1 & 3.2$\pm$1.1 & \textbf{25.3}$\pm$4.5 & 24.1$\pm$4.2 \\
        & 0.1\%  & 87.4$\pm$4.2 & \textbf{95.5}$\pm$0.9 & 94.1$\pm$1.1 & 90.2$\pm$5.2 & \textbf{98.4}$\pm$0.6 & 97.7$\pm$0.8 & 7.8$\pm$2.4 & \textbf{55.4}$\pm$6.1 & 54.2$\pm$6.3 \\
        & 0.5\%  & 89.7$\pm$5.5 & \textbf{97.2}$\pm$0.6 & 96.3$\pm$0.8 & 93.4$\pm$4.8 & \textbf{99.7}$\pm$0.2 & 99.1$\pm$0.3 & 15.6$\pm$4.1 & \textbf{89.1}$\pm$2.3 & 88.5$\pm$2.5 \\
        & 2.0\%  & 94.3$\pm$1.7 & \textbf{99.1}$\pm$0.3 & 98.5$\pm$0.4 & 96.8$\pm$1.5 & \textbf{100.0}$\pm$0.0 & 99.8$\pm$0.1 & 52.4$\pm$6.7 & \textbf{97.6}$\pm$1.0 & 96.4$\pm$1.2 \\
        & 5.0\%  & 97.5$\pm$0.9 & \textbf{99.8}$\pm$0.1 & 99.6$\pm$0.2 & 98.9$\pm$0.4 & \textbf{100.0}$\pm$0.0 & \textbf{100.0}$\pm$0.0 & 88.2$\pm$2.6 & \textbf{99.5}$\pm$0.3 & 98.9$\pm$0.4 \\
        \midrule
        \multirow{6}{*}{ViMoE} 
        & 0.01\% & 2.5$\pm$0.5 & \textbf{6.4}$\pm$1.5 & 5.7$\pm$1.3 & 8.4$\pm$0.8 & \textbf{12.1}$\pm$2.5 & 11.4$\pm$2.2 & 0.8$\pm$0.3 & \textbf{2.5}$\pm$0.7 & 2.0$\pm$0.6 \\
        & 0.05\% & 15.6$\pm$1.1 & \textbf{88.3}$\pm$2.4 & 87.1$\pm$2.6 & 28.5$\pm$1.7 & \textbf{92.4}$\pm$1.7 & 91.5$\pm$1.9 & 1.5$\pm$0.5 & \textbf{18.2}$\pm$3.6 & 17.1$\pm$3.4 \\
        & 0.1\%  & 45.2$\pm$3.2 & \textbf{95.4}$\pm$1.1 & 94.2$\pm$1.4 & 62.1$\pm$4.1 & \textbf{97.8}$\pm$0.8 & 96.9$\pm$1.1 & 4.2$\pm$1.2 & \textbf{48.6}$\pm$5.8 & 47.4$\pm$6.0 \\
        & 0.5\%  & 88.9$\pm$5.8 & \textbf{99.1}$\pm$0.4 & 98.4$\pm$0.6 & 94.3$\pm$6.3 & \textbf{99.8}$\pm$0.1 & 99.4$\pm$0.3 & 12.5$\pm$3.1 & \textbf{85.7}$\pm$2.7 & 84.5$\pm$2.9 \\
        & 2.0\%  & 95.8$\pm$2.1 & \textbf{100.0}$\pm$0.0 & 99.7$\pm$0.2 & 98.2$\pm$1.8 & \textbf{100.0}$\pm$0.0 & \textbf{100.0}$\pm$0.0 & 45.3$\pm$6.2 & \textbf{98.1}$\pm$0.8 & 97.2$\pm$1.1 \\
        & 5.0\%  & 98.1$\pm$0.7 & \textbf{100.0}$\pm$0.0 & \textbf{100.0}$\pm$0.0 & 99.6$\pm$0.2 & \textbf{100.0}$\pm$0.0 & \textbf{100.0}$\pm$0.0 & 84.5$\pm$3.1 & \textbf{99.6}$\pm$0.2 & 99.1$\pm$0.4 \\
        \midrule
        \multirow{6}{*}{pMoE} 
        & 0.01\% & 8.7$\pm$1.4 & \textbf{10.5}$\pm$2.1 & 9.8$\pm$1.9 & 3.8$\pm$1.1 & \textbf{83.2}$\pm$3.5 & 82.4$\pm$3.8 & 4.6$\pm$1.2 & \textbf{6.8}$\pm$1.5 & 5.3$\pm$1.2 \\
        & 0.05\% & 58.5$\pm$3.1 & \textbf{96.8}$\pm$1.2 & 96.6$\pm$1.3 & 12.8$\pm$2.6 & \textbf{96.5}$\pm$1.1 & 95.3$\pm$1.3 & 6.3$\pm$1.6 & \textbf{13.5}$\pm$2.8 & 11.8$\pm$2.5 \\
        & 0.1\%  & 76.3$\pm$5.6 & \textbf{98.5}$\pm$0.5 & 98.3$\pm$0.6 & 42.3$\pm$5.4 & \textbf{99.1}$\pm$0.4 & 98.3$\pm$0.6 & 9.1$\pm$2.1 & \textbf{21.0}$\pm$4.3 & 19.1$\pm$4.5 \\
        & 0.5\%  & 95.2$\pm6.2$ & \textbf{99.9}$\pm$0.1 & 99.8$\pm$0.1 & 94.5$\pm$4.8 & \textbf{100.0}$\pm$0.0 & 99.8$\pm$0.1 & 18.7$\pm$4.2 & \textbf{76.5}$\pm$4.1 & 74.9$\pm$4.4 \\
        & 2.0\%  & 99.7$\pm$1.7 & \textbf{100.0}$\pm$0.0 & \textbf{100.0}$\pm$0.0 & 99.5$\pm$1.2 & \textbf{100.0}$\pm$0.0 & \textbf{100.0}$\pm$0.0 & 64.5$\pm$5.8 & \textbf{96.8}$\pm$1.2 & 95.6$\pm$1.5 \\
        & 5.0\%  & 99.9$\pm$0.1 & \textbf{100.0}$\pm$0.0 & \textbf{100.0}$\pm$0.0 & 99.9$\pm$0.1 & \textbf{100.0}$\pm$0.0 & \textbf{100.0}$\pm$0.0 & 91.2$\pm$2.1 & \textbf{99.5}$\pm$0.3 & 98.8$\pm$0.5 \\
        \bottomrule
    \end{tabular}
\end{table*}

\textbf{Superior Performance at Low Poisoning Rates.} Both variants of \ourmethod achieve high ASR at poisoning rates where the routing-agnostic approach fails. For instance, on the V-MoE architecture at a 0.05\% poisoning rate, the routing-agnostic square trigger manages 1.6\% ASR. In contrast, \ourmethodnospace-S achieves 84.5\% ASR, and \ourmethodnospace-N reaches 86.1\%. This demonstrates that aligning the trigger with the router's patch selection mechanism is necessary for low-budget backdoor viability in large-parameter vision MoE models.

\textbf{Higher Peak ASR.} Across all datasets and model capacities, both \ourmethod formulations hit near-perfect saturation ($\ge$ 99\%) with lower poisoning rates than routing-agnostic triggers. While the routing-agnostic warped trigger on ViMoE only reaches 89.5\% ASR even at a 5.0\% poisoning rate, \ourmethod achieves $\ge$ 99.1\% ASR under the same conditions. Ensuring that active experts are consistently forced to process the complete trigger pattern eliminates the fragmentation issues that previously bottlenecked peak performance.

\textbf{The Naive vs. Saliency-Guided Trade-Off.} \ourmethodnospace-N consistently achieves the highest absolute ASR, outperforming \ourmethodnospace-S. However, this margin is extremely narrow and never exceeds a 2\% difference across all recorded experiments. Because a naive application overwrites every conceptual patch in the image, it guarantees 100\% routing saturation but incurs severe visual distortion. The saliency-guided approach strategically targets only the dominant foreground patches, achieving nearly identical attack performance while structurally preserving the benign visual characteristics of the background.

\begin{figure}[!t]
    \centering
    \includegraphics{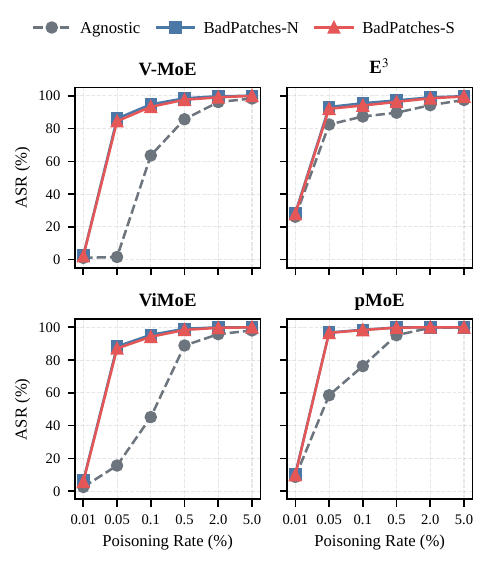}
    \caption{ASR plotted against poisoning rate.}
    \label{fig:asr_curve}
\end{figure}

To illustrate the efficiency of \ourmethod in low-budget scenarios, Figure~\ref{fig:asr_curve} plots the ASR as a function of the poisoning rate. As shown, \ourmethodnospace-S (solid lines) displays an upward trajectory, reaching high-performance thresholds while the routing-agnostic baseline (dashed lines) remains dormant. This figure highlights the stark contrast in attack viability, particularly at poisoning rates below 0.1\%, where our saliency-guided approach achieves significant performance gains compared to the stagnant baseline.

Overall, the data confirms that optimizing trigger placement for spatial routing shows unique vulnerabilities in sparse MoE vision models, even as parameter counts scale up significantly.

\subsection{Impact on Model Utility}
\label{sec:utility_impact}

To ensure the stealthiness of the backdoor, it is important that the attack does not significantly degrade the model's performance on clean inputs. Table~\ref{tab:utility_impact} reports the Benign Accuracy (BA) and Clean Accuracy Drop (CAD) of models poisoned with \ourmethodnospace-S compared to the routing-agnostic baselines at the specific poisoning rate that achieved the maximum ASR for each configuration.

\begin{table}[!t]
    \centering
    \scriptsize
    \caption{Impact on model utility at peak attack performance. We report the BA\% and CAD\% strictly for the poisoning rate that achieved the highest ASR for each specific trigger and method combination.}
    \label{tab:utility_impact}
    \begin{tabular}{ll | cc cc}
        \toprule
        \multirow{2}{*}{\textbf{Model}} & \multirow{2}{*}{\textbf{Trigger}} & \multicolumn{2}{c}{\textbf{Agnostic}} & \multicolumn{2}{c}{\textbf{\ourmethodnospace-S}} \\
        \cmidrule(lr){3-4} \cmidrule(lr){5-6}
        & & \textbf{BA $\uparrow$} & \textbf{CAD $\downarrow$} & \textbf{BA $\uparrow$} & \textbf{CAD $\downarrow$} \\
        \midrule
        \multirow{3}{*}{V-MoE} 
        & Square & 91.3$\pm$0.4 & 1.8$\pm$0.3 & \textbf{91.5}$\pm$0.3 & \textbf{1.6}$\pm$0.2 \\
        & Blend  & \textbf{90.7}$\pm$0.5 & \textbf{2.4}$\pm$0.4 & 90.5$\pm$0.4 & 2.6$\pm$0.3 \\
        & Warped & \textbf{92.4}$\pm$0.3 & \textbf{0.7}$\pm$0.2 & 92.1$\pm$0.3 & 1.0$\pm$0.2 \\
        \midrule
        \multirow{3}{*}{E$^3$} 
        & Square & 91.5$\pm$0.6 & 2.3$\pm$0.4 & \textbf{91.8}$\pm$0.5 & \textbf{2.0}$\pm$0.3 \\
        & Blend  & \textbf{90.9}$\pm$0.7 & \textbf{2.9}$\pm$0.5 & 90.7$\pm$0.5 & 3.1$\pm$0.4 \\
        & Warped & \textbf{93.4}$\pm$0.4 & \textbf{0.4}$\pm$0.1 & 93.1$\pm$0.4 & 0.7$\pm$0.2 \\
        \midrule
        \multirow{3}{*}{ViMoE} 
        & Square & 86.4$\pm$0.5 & 1.6$\pm$0.3 & \textbf{86.8}$\pm$0.4 & \textbf{1.2}$\pm$0.2 \\
        & Blend  & 85.9$\pm$0.6 & 2.1$\pm$0.4 & \textbf{86.1}$\pm$0.5 & \textbf{1.9}$\pm$0.3 \\
        & Warped & \textbf{87.1}$\pm$0.4 & \textbf{0.9}$\pm$0.2 & 87.0$\pm$0.4 & 1.0$\pm$0.2 \\
        \midrule
        \multirow{3}{*}{pMoE} 
        & Square & 88.4$\pm$0.4 & 1.6$\pm$0.3 & \textbf{88.8}$\pm$0.3 & \textbf{1.2}$\pm$0.2 \\
        & Blend  & 88.5$\pm$0.5 & 1.5$\pm$0.3 & \textbf{88.6}$\pm$0.4 & \textbf{1.4}$\pm$0.2 \\
        & Warped & \textbf{88.9}$\pm$0.3 & \textbf{1.1}$\pm$0.2 & 88.8$\pm$0.3 & 1.2$\pm$0.2 \\
        \bottomrule
    \end{tabular}
\end{table}

The results show there is no large negative impact on the normal accuracy of the poisoned models. Comparing the BA and CAD between all experiments, the performance of models backdoored using \ourmethodnospace-S aligns closely with the routing-agnostic baselines, consistently remaining within a 0.3\% margin and frequently outperforming the baseline on the localized square trigger. This indicates that strategically localizing the trigger to dominant foreground patches successfully maintains the model's ability to learn normal semantic representations without sacrificing peak attack performance.

\subsection{Impact of Varying the Number of Patches}
\label{subsec:varying_patches}

In this subsection, we investigate how the hyperparameter $l$, the total number of conceptual patches an image is divided into, affects the performance of \ourmethodnospace. We run this evaluation exclusively on the V-MoE architecture using all three trigger types (Square, Blend, and Warped) at a fixed poisoning rate of 0.1\%. We select a poisoning rate of 0.1\% because it is sufficient to yield partial success for the routing-agnostic baseline without reaching ASR saturation. In our main experiments, V-MoE divides the $384 \times 384$ input images into $l=576$ patches (a $24 \times 24$ grid). Here, we evaluate the attack across lower values of $l \in \{256, 144, 64, 16\}$. Decreasing $l$ inherently increases the spatial dimensions of each patch. However, the router's capacity constraint (top-$n$ patches) scales proportionally, meaning the model consistently discards 50\% of the image patches regardless of the grid resolution.

Table~\ref{tab:varying_patches} presents the ASR for these configurations. From these results, we derive two key observations regarding the relationship between patch size and routing-aware triggers.

\begin{table*}[tb]
    \centering
    \scriptsize
    \caption{Impact of varying the total number of patches ($l$) on V-MoE at a 0.1\% poisoning rate. \ourmethodnospace-S remains effective and consistent across all resolutions, while the routing-agnostic baselines struggle at higher grid densities due to patch fragmentation.}
    \label{tab:varying_patches}
    \begin{tabular}{c | cc cc cc}
        \toprule
        \multirow{2}{*}{\textbf{Patches ($l$)}} & \multicolumn{2}{c}{\textbf{Square (ASR $\uparrow$)}} & \multicolumn{2}{c}{\textbf{Blend (ASR $\uparrow$)}} & \multicolumn{2}{c}{\textbf{Warped (ASR $\uparrow$)}} \\
        \cmidrule(lr){2-3} \cmidrule(lr){4-5} \cmidrule(lr){6-7}
        & \textbf{Agnostic} & \textbf{\ourmethodnospace-S} & \textbf{Agnostic} & \textbf{\ourmethodnospace-S} & \textbf{Agnostic} & \textbf{\ourmethodnospace-S} \\
        \midrule
        576 & 63.6$\pm$1.1 & \textbf{93.3}$\pm$1.2 & 70.1$\pm$3.4 & \textbf{96.5}$\pm$0.9 & 2.4$\pm$0.6  & \textbf{41.3}$\pm$5.5 \\
       256        & 68.2$\pm$2.3 & \textbf{94.1}$\pm$1.0 & 74.5$\pm$2.8 & \textbf{96.8}$\pm$1.1 & 5.8$\pm$1.4  & \textbf{41.8}$\pm$4.2 \\
        144        & 75.4$\pm$1.9 & \textbf{93.8}$\pm$1.1 & 80.2$\pm$3.1 & \textbf{96.4}$\pm$0.8 & 10.4$\pm$2.1 & \textbf{41.1}$\pm$5.1 \\
        64         & 81.0$\pm$2.5 & \textbf{94.0}$\pm$0.9 & 86.7$\pm$2.5 & \textbf{96.7}$\pm$1.0 & 16.2$\pm$2.7 & \textbf{42.0}$\pm$4.8 \\
        16         & 81.5$\pm$2.1 & \textbf{93.9}$\pm$1.0 & 90.4$\pm$1.9 & \textbf{96.6}$\pm$1.2 & 23.5$\pm$3.2 & \textbf{41.5}$\pm$5.3 \\
        \bottomrule
    \end{tabular}
\end{table*}

\textbf{Consistent Superiority of Saliency-Guided Targeting.} Regardless of the patch grid resolution, \ourmethodnospace-S maintains a consistent ASR across all trigger types (e.g., holding steady at $\ge$ 93.3\% for the Square trigger), proving that the attack's performance is robust against variations in the model's patch extraction configuration. By explicitly injecting the trigger payload into the most salient patches, \ourmethod successfully forces the malicious pattern through the routing bottleneck, entirely bypassing the 50\% patch discarding mechanism.

\textbf{Agnostic Performance Scales Inversely with $l$.} As the number of patches decreases (resulting in larger patch sizes), the ASR of the routing-agnostic baselines notably improves. For example, the agnostic Square trigger ASR rises from 63.6\% at $l=576$ to approximately 81.5\% at $l=16$, with similar inverse scaling trends observed for the Blend and Warped triggers.

The behavior of the routing-agnostic triggers is governed by two interacting factors: spatial fragmentation and patch saliency. When $l$ is high (e.g., 576), patches are small, causing statically placed or globally applied agnostic triggers to be fragmented across multiple patch boundaries. Furthermore, a small background patch containing a piece of the trigger often lacks sufficient semantic features to achieve a high routing score, causing the router to discard it entirely. As patches grow larger at lower values of $l$, this fragmentation risk decreases, allowing the complete agnostic pattern to be encapsulated within a single patch.

As patches grow larger at lower values of $l$, both issues are mitigated. First, the fragmentation risk decreases, allowing larger portions of the agnostic pattern to be encapsulated within a single patch. More importantly, large patches cover a significant portion of the image. An agnostic trigger is now highly likely to share a patch with actual foreground elements (e.g., the edge of a salient object). Because the router evaluates the entire patch, the trigger inadvertently ``hitchhikes'' on the natural saliency of these foreground features, increasing its probability of being selected by the active experts.

Despite this improvement at lower grid resolutions, the agnostic baseline still underperforms compared to \ourmethodnospace-S. Because the router drops 50\% of the patches based on content saliency, a statically placed agnostic trigger remains susceptible to being discarded or broken up if the patch containing the trigger does not contain enough salient features to be routed to an expert. Consequently, the success of the agnostic baseline relies more on accidental foreground overlap. In contrast, \ourmethod identifies and alters the patches the router is more likely to select, ensuring the backdoor is learned more efficiently at any resolution.

\subsection{Gray-Box Analysis of Unknown Patch Configurations}
\label{subsec:gray_box_analysis}

\ourmethod explicitly focuses on data poisoning. Because our saliency-guided method relies purely on input image features and data poisoning rather than internal routing states, it is suited for restricted access scenarios. However, in a realistic gray-box environment, an attacker may know the target uses a vision MoE architecture but lacks exact knowledge of its patching configuration, e.g., the setting for hyperparameter $l$. To determine the attack's viability under these conditions, we evaluate \ourmethodnospace-S on V-MoE, continuing with the isolated 0.1\% poisoning rate established in Section~\ref{subsec:varying_patches}. In these experiments, the attacker generates the poisoned dataset assuming a specific grid resolution ($l_{attacker}$), while the model is trained using a different configuration ($l_{model}$).

Table~\ref{tab:gray_box} presents the ASR when $l_{model}$ and $l_{attacker}$ are mismatched. The results demonstrate that precise knowledge of the patch configuration is required for optimal performance. However, \ourmethod remains highly effective even with incorrect assumptions, consistently outperforming the routing-agnostic baselines. The performance variance depends on the spatial relationship between the assumed and actual grids.

\begin{table}[!t]
    \centering
    \scriptsize
    \caption{Experimental results of \ourmethodnospace-S on V-MoE at a 0.1\% poisoning rate under gray-box conditions. The attacker assumes a grid size ($l_{attacker}$) that differs from the actual model configuration ($l_{model}$). The attack remains robust, though slight fragmentation penalties occur when $l_{model} > l_{attacker}$.}
    \label{tab:gray_box}
    \begin{tabular}{cc | ccc}
        \toprule
        \multirow{2}{*}{\textbf{$l_{model}$}} & \multirow{2}{*}{\textbf{$l_{attacker}$}} & \textbf{Square} & \textbf{Blend} & \textbf{Warped} \\
        & & \textbf{(ASR $\uparrow$)} & \textbf{(ASR $\uparrow$)} & \textbf{(ASR $\uparrow$)} \\
        \midrule
        \multirow{2}{*}{576} 
        & 256 & 90.1$\pm$1.3 & 93.2$\pm$1.1 & 38.5$\pm$4.6 \\
        & 64  & 88.4$\pm$1.5 & 91.5$\pm$1.4 & 35.1$\pm$5.2 \\
        \midrule
        \multirow{2}{*}{256} 
        & 576 & 93.8$\pm$1.1 & 96.6$\pm$1.0 & 41.5$\pm$4.1 \\
        & 64  & 89.2$\pm$1.4 & 92.4$\pm$1.2 & 37.2$\pm$4.8 \\
        \midrule
        \multirow{2}{*}{64} 
        & 576 & 94.0$\pm$0.9 & 96.7$\pm$0.8 & 41.8$\pm$4.0 \\
        & 256 & 93.9$\pm$1.0 & 96.5$\pm$0.9 & 41.6$\pm$4.2 \\
        \bottomrule
    \end{tabular}
\end{table}

\textbf{The Encapsulation Effect ($l_{model} < l_{attacker}$).} When the model uses a coarser grid than the attacker assumed, the complete processing of the trigger is guaranteed. Because the attacker generates a trigger sized for a smaller patch, the model's larger patch completely encapsulates the malicious pattern. This preserves the trigger's structural integrity, allowing the ASR to remain near its peak (e.g., $\ge$ 93.8\% for the Square trigger).

\textbf{The Fragmentation Effect ($l_{model} > l_{attacker}$).} When the model uses a finer grid than the attacker anticipated, the attack performance slightly decreases. An attacker who assumes large patches will generate a correspondingly large trigger. When the model subsequently divides the image into smaller patches, the trigger is sliced across multiple patch boundaries. This fragmentation makes it more difficult for the active experts to reliably associate the trigger with the target label, causing a minor ASR drop (e.g., decreasing to 88.4\% for the Square trigger when $l_{model}=576$ and $l_{attacker}=64$).

Despite the fragmentation, the attack does not fail. Because the attacker still places the trigger in the highly salient regions of the image, the router is more likely to select the fragmented pieces. This ensures the backdoor is still learned more efficiently than an arbitrarily placed agnostic trigger.

\begin{table*}[!t]
    \centering
    \scriptsize
    \caption{ASR \% of V-MoE at a 0.1\% poisoning rate evaluated on asymmetric patch grids. \ourmethodnospace-S maintains peak performance regardless of whether the resulting patches are square or rectangular, proving the attack does not rely on symmetric image cropping.}
    \label{tab:asymmetric_patches}
    \begin{tabular}{c c c | cc cc cc}
        \toprule
        \multirow{2}{*}{\textbf{Grid Layout}} & \multirow{2}{*}{\textbf{Total Patches ($l$)}} & \multirow{2}{*}{\textbf{Patch Dims (px)}} & \multicolumn{2}{c}{\textbf{Square (ASR $\uparrow$)}} & \multicolumn{2}{c}{\textbf{Blend (ASR $\uparrow$)}} & \multicolumn{2}{c}{\textbf{Warped (ASR $\uparrow$)}} \\
        \cmidrule(lr){4-5} \cmidrule(lr){6-7} \cmidrule(lr){8-9}
        & & & \textbf{Agnostic} & \textbf{\ourmethodnospace-S} & \textbf{Agnostic} & \textbf{\ourmethodnospace-S} & \textbf{Agnostic} & \textbf{\ourmethodnospace-S} \\
        \midrule
        24 $\times$ 24 & 576 & 16 $\times$ 16 & 63.6$\pm$1.1 & \textbf{93.3}$\pm$1.2 & 70.1$\pm$3.4 & \textbf{96.5}$\pm$0.9 & 2.4$\pm$0.6  & \textbf{41.3}$\pm$5.5 \\
        16 $\times$ 24        & 384 & 24 $\times$ 16 & 67.5$\pm$1.8 & \textbf{94.5}$\pm$1.1 & 73.8$\pm$2.8 & \textbf{96.3}$\pm$1.0 & 4.2$\pm$1.2  & \textbf{41.6}$\pm$4.9 \\
        12 $\times$ 16        & 192 & 32 $\times$ 24 & 71.4$\pm$2.1 & \textbf{94.8}$\pm$1.0 & 76.8$\pm$2.6 & \textbf{96.2}$\pm$1.1 & 6.2$\pm$1.5  & \textbf{42.9}$\pm$4.3 \\
        8 $\times$ 12         & 96  & 48 $\times$ 32 & 78.5$\pm$2.4 & \textbf{95.1}$\pm$0.9 & 83.1$\pm$2.1 & \textbf{96.7}$\pm$1.0 & 12.1$\pm$2.4 & \textbf{43.8}$\pm$4.5 \\
        \bottomrule
    \end{tabular}
\end{table*}

\subsection{Robustness to Asymmetric Patch Configurations}
\label{subsec:asymmetric_patches}

Standard vision MoE evaluations often assume square input images symmetrically divided into square patches, e.g., a $24 \times 24$ grid. However, real-world datasets frequently contain images with varying aspect ratios, such as 16:9 or 4:3. Processing these images may be done via asymmetric grid divisions, yielding grid shapes such as $8 \times 12$ or $9 \times 16$, which result in non-square, rectangular patches. To address whether our methodology relies on symmetric square cropping, we evaluate the robustness of \ourmethodnospace-S against arbitrary asymmetric patch configurations.

We run this evaluation on the V-MoE architecture, again at a fixed 0.1\% poisoning rate. We restructure the patch extraction layer to process inputs using varying asymmetric grids while maintaining the standard 50\% patch-discarding rate. Table~\ref{tab:asymmetric_patches} presents the ASR across these non-standard configurations. From these results, we observe that \ourmethod remains highly effective regardless of the patch geometry.

\textbf{Algorithmic Flexibility to Patch Shape.} \ourmethodnospace-S maintains a consistently high ASR across all asymmetric grids, matching the performance observed in symmetric configurations. This robustness stems from the design of our saliency scoring function. Because the Spectral Residual algorithm calculates saliency at the pixel level before aggregating the scores within the spatial boundaries of each patch, the geometric shape of that boundary (square versus rectangular) does not impede the algorithm's ability to locate dominant foreground elements.

\textbf{Agnostic Degradation under Geometric Shifts.} The routing-agnostic baselines experience unpredictable fluctuations with asymmetric grids. A static square trigger placed in the corner of a 16:9 image divided into rectangular patches suffers from irregular geometric fragmentation. Depending on the exact grid division, the static trigger may be sliced awkwardly across boundary lines, further demonstrating the fragility of applying routing-agnostic triggers to sparse, patch-based models.

Ultimately, the MoE router evaluates and routes tokens based on the semantic density of their encapsulated pixels, regardless of patch shape. By dynamically injecting the trigger into the patches with the highest semantic density, \ourmethodnospace-S improves routing saturation and successful backdoor injection independent of the image's aspect ratio or the cropping geometry.

\subsection{Defense Analysis}
\label{sec:defense_results}

To evaluate the robustness of \ourmethodnospace-S against standard mitigation strategies, we consider the fine-pruning defense framework~\cite{liu2018finepruning}. Fine-pruning is a widely adopted defense for non-MoE architectures~\cite{hong2022handcraftedbackdoors,nguyen2021wanet} that operates in two distinct phases: first, it prunes dormant neurons based on clean data activations, and second, it performs a brief fine-tuning phase to recover any lost accuracy.

We evaluate this defense on a backdoored V-MoE model trained with the \ourmethodnospace-S square trigger at a 0.1\% poisoning rate. Because V-MoE utilizes Transformer blocks rather than convolutions, we adapt the defense to prune the neurons within the MLP layers of the model's experts. We test pruning rates of 10\%, 20\%, and 30\%, followed by a limited fine-tuning phase of 5 epochs to simulate a realistic computational budget constraint. All reported results are averaged over three independent runs.

Table~\ref{tab:fine_pruning} presents the BA and ASR after each phase of the defense. From these results, we derive two observations regarding the inadequacy of traditional defenses when applied to vision MoEs.

\begin{table}[!t]
    \centering
    \scriptsize
    \caption{Results of the fine-pruning defense framework evaluated on backdoored V-MoE models. The framework is split into the Pruning phase and the Fine-Tuning phase. A higher ASR indicates the backdoor attack is more resilient against the defense.}
    \label{tab:fine_pruning}
    \begin{tabular}{c l | cc cc}
        \toprule
        \multirow{2}{*}{\textbf{Rate}} & \multirow{2}{*}{\textbf{Phase}} & \multicolumn{2}{c}{\textbf{Agnostic}} & \multicolumn{2}{c}{\textbf{\ourmethodnospace-S}}\\
        \cmidrule(lr){3-4}\cmidrule(lr){5-6}
        & & \textbf{BA $\uparrow$} & \textbf{ASR $\uparrow$} & \textbf{BA $\uparrow$} & \textbf{ASR $\uparrow$}\\
        \midrule
        & Initial     & \textbf{92.0}$\pm$0.7 & 63.6$\pm$1.1 & 91.7$\pm$0.8 & \textbf{93.3}$\pm$1.2 \\
        \midrule
        \multirow{2}{*}{10\%} & Pruning     & \textbf{88.4}$\pm$0.6 & 61.2$\pm$1.5 & 88.1$\pm$0.5 & \textbf{92.8}$\pm$0.9 \\
                              & Fine-Tuning & \textbf{91.5}$\pm$0.5 & 4.5$\pm$0.8  & 91.1$\pm$0.4 & \textbf{35.2}$\pm$1.8 \\
        \midrule
        \multirow{2}{*}{20\%} & Pruning     & \textbf{83.2}$\pm$0.8 & 62.5$\pm$1.2 & 82.5$\pm$0.7 & \textbf{92.4}$\pm$1.0 \\
                              & Fine-Tuning & \textbf{91.1}$\pm$0.6 & 3.8$\pm$0.7  & 90.5$\pm$0.5 & \textbf{33.6}$\pm$2.1 \\
        \midrule
        \multirow{2}{*}{30\%} & Pruning     & \textbf{74.5}$\pm$1.2 & 62.1$\pm$1.4 & 73.8$\pm$1.1 & \textbf{91.5}$\pm$1.3 \\
                              & Fine-Tuning & \textbf{90.8}$\pm$0.7 & 4.1$\pm$1.1  & 90.1$\pm$0.6 & \textbf{31.4}$\pm$2.5 \\
        \bottomrule
    \end{tabular}
\end{table}

\textbf{Ineffectiveness of Activation-Based Pruning.} The results demonstrate that the pruning phase alone fails to remove the backdoor. Across all pruning rates, the ASR drop is a maximum of 2.4\% for both the routing-agnostic baseline and \ourmethodnospace-S immediately after pruning. This contrasts with results on dense models in prior work and the premise of the fine-pruning defense~\cite{liu2018finepruning}, which assumes that the backdoor is isolated in specific dormant neurons that can be pruned away. Because \ourmethodnospace-S targets salient foreground patches, the adversarial payload is routed to the same experts as important benign features. Consequently, the backdoor representations become entangled with normal semantic features inside the expert MLPs, making standard activation-based pruning ineffective.

\textbf{The Cost of Fine-Tuning Recovery.} While the complete fine-pruning framework (pruning followed by fine-tuning) mitigates the routing-agnostic backdoor, reducing the ASR to 3.8\%, it is less effective against \ourmethodnospace-S, which maintains an ASR of 31.4\%. Furthermore, relying on fine-tuning to scrub backdoors is impractical for MoE models. As shown in Table~\ref{tab:fine_pruning}, increasing the pruning rate causes drops in BA (e.g., dropping to 73.8\% before fine-tuning). Recovering this accuracy requires fine-tuning. Given the massive parameter counts of architectures like V-MoE (14.7B total parameters), the computational expense of this fine-tuning phase presents a trade-off between removing the backdoor and the cost of doing so.

\begin{figure}[!t]
    \centering
    \includegraphics{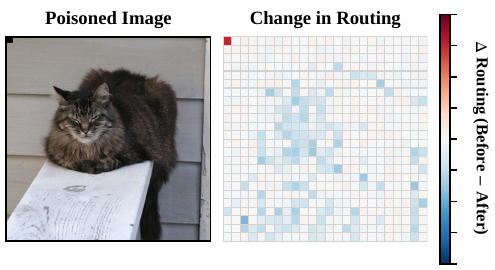}
    \caption{Routing heatmaps for a V-MoE model processing an image poisoned with a routing-agnostic square trigger. Left: The original poisoned image. Right: Heatmap showing the difference in routing before and after the full fine-pruning defense. The red color indicates that a patch was processed more often before the fine-pruning defense, showing that the patch containing the trigger is processed less often after the defense is applied.}
    \label{fig:patch_routing_pruning}
\end{figure}

Figure~\ref{fig:patch_routing_pruning} visualizes the effect of the fine-pruning framework on the patch routing. Before the defense is applied, the experts in the model consistently process the patch containing the square trigger, indicating the trigger's influence on the classification output. After the complete fine-pruning process, the model reallocates its routing probabilities, processing the trigger patch fewer times. This visualizes why the attack's effectiveness decreases: the fine-tuning on clean images and labels unlearns the trigger-target association.

\section{Discussion}

\textbf{Security Implications for Vision MoEs.} The fundamental appeal of Vision MoE architectures lies in their computational efficiency; by discarding uninformative patches and conditionally activating parameters, these models can scale to billions of parameters without incurring prohibitive inference costs. However, our findings highlight that this exact sparsity mechanism amplifies the model's vulnerability to data poisoning. \ourmethod exploits the routing mechanism's bias toward foreground saliency. By applying the adversarial payload within the patches, the model is structurally more likely to process, and an attacker can effectively hijack the dominant routing paths. 

This presents a supply-chain threat. Because \ourmethodnospace-S achieves near-perfect attack success rates at poisoning budgets as low as 0.01\% to 0.05\%, the attack can be considered more stealthy. As the industry increasingly relies on massive, uncurated open-source datasets and datasets obtained via web scraping to train foundational MoE models, the feasibility of an adversary strategically poisoning a few datasets presents a vulnerability. 

\textbf{Defending against Routing-Aware Backdoors.} The failure of standard backdoor mitigation strategies, such as pruning, highlights a need for novel, routing-aware defenses. As demonstrated by our evaluation of the fine-pruning framework, activation-based pruning fails because \ourmethodnospace-S targets salient foreground patches. Consequently, the backdoor representations become entangled with benign features inside the expert MLPs, making it more difficult to prune the malicious neurons without reducing the model's primary utility.

Securing sparse models will likely require a shift from analyzing static neuron activations to monitoring dynamic routing behaviors. For example, defenders could analyze expert load balancing during inference; if a specific expert exhibits abnormally high selection frequencies exclusively for a particular predicted class, it may indicate a compromised routing pathway. Alternatively, input-level preprocessing defenses, such as randomized patch dropping, spatial cropping, or patch shuffling during inference, could be explored to disrupt the spatial integrity of the salient patches, preventing the complete trigger from reaching the active experts.

\textbf{Limitations of \ourmethodnospace.} While our saliency-guided approach significantly improves attack efficacy, our current evaluation operates under a dirty-label threat model, where the adversary possesses the capability to alter both the image and its corresponding training label. While this aligns with supply-chain data poisoning scenarios, it may limit the attack's applicability in highly regulated environments where labels are strictly audited, but images are not. Adapting the \ourmethod routing-aware framework for clean-label backdoor attacks presents a promising direction for future research, as it would require balancing the trigger's routing saliency against the semantic integrity of the original class.
\section{Related Work}
\label{sec:related_work}

Backdoor attacks, or data poisoning, refer to adversarial techniques that embed predefined triggers into training samples to manipulate model behavior during inference~\cite{gu2019badnets}. These attacks highlight systematic failure modes in model training: models may learn spurious correlations between the hidden trigger and a target label, compromising their integrity when deployed on malicious inputs~\cite{gu2019badnets}. 

Prior work spans localized, blended, and stealthy trigger designs for dense architectures. Backdoor attacks were originally introduced by Gu et al.~\cite{gu2019badnets} with BadNets, a method that embeds localized, fixed-pattern triggers like colored squares into training samples to successfully induce misclassification. Building on this foundation, other work introduced trigger design and application methods that make triggers more robust and harder to remove. For example, Chen et al.~\cite{chen2017blend} proposed blended triggers that overlay adversarial images onto benign samples to increase attack robustness. However, these trigger designs are still easily detected via human inspection. Because of this, a major line of subsequent backdoor attack research focused on the direction of visual imperceptibility. Nguyen et al.~\cite{nguyen2021wanet} introduced warped triggers, which geometrically displace pixels to evade detection by human moderators. While these methods are highly effective on dense CNNs and standard vision transformers, they are fundamentally routing-agnostic. They assume uniform processing of the input space, an assumption that breaks down in sparse, conditional architectures where input images are fragmented, and pieces are conditionally dropped.

While MoE architectures are becoming prevalent, their security against backdoor attacks remains underexplored, particularly in the vision domain. Initial MoE security research has primarily focused on language models, revealing that the sparse expert routing mechanism can itself become an attack surface~\cite{wu2025gatebreaker,lintelo2026largelanguagelobotomy}. Recent works demonstrate that adversaries can compromise model alignment and performance by injecting text-based triggers. For instance, Wang et al.~\cite{wang2025badmoe} showed that MoE language models can be manipulated to misclassify sentiment analysis tasks. Similarly, Zhang et al.~\cite{zhang2024instructionbackdoor} demonstrated successful text instruction backdoor attacks against MoE models. However, these interventions operate primarily at the token or phrase level in the natural language processing domain and do not account for the localized patch-routing mechanisms present in vision-based MoE models.

\textbf{Distinction from Prior Methods.} While existing backdoor methodologies demonstrate the vulnerability of dense vision models, they fundamentally rely on a routing-agnostic application of triggers, or the MoE-specific methodologies are only tested on MoE LLMs. In contrast, \ourmethod demonstrates that ignoring the router's selective patch-dropping mechanism leads to brittle and inefficient attacks in sparse vision architectures. By applying the complete trigger to individual conceptual patches, our approach actively accounts for the dynamic patch-processing inherent to vision MoEs. This allows \ourmethod to ensure the complete trigger is processed by the experts regardless of the specific routing configuration. By isolating these structural routing dependencies rather than relying on uniform full-image processing, \ourmethod achieves a significantly higher attack success rate while preserving general model utility at low poisoning rates.
\section{Conclusion and Future Work}
\label{sec:conclusion}

In this work, we investigated the security vulnerabilities introduced by the patch-based processing mechanisms of vision Mixture-of-Experts (MoE) architectures. While conditional computation and patch discarding significantly reduce inference costs, they also disrupt traditional, routing-agnostic backdoor attacks by fragmenting or entirely discarding adversarial patterns. To show this architectural vulnerability, we introduced \ourmethodnospace, a novel routing-aware trigger application strategy. By utilizing a saliency-guided approach, \ourmethod ensures that adversarial triggers are fully encapsulated within dominant foreground patches, guaranteeing they are consistently processed by the active experts rather than being discarded by the router.

Our evaluations across four state-of-the-art vision MoE models (V-MoE, $E^3$, ViMoE, and pMoE) demonstrate that aligning the attack with the model's spatial routing dynamics improves backdoor effectiveness. \ourmethod achieves a high Attack Success Rate (ASR) at low poisoning rates, while preserving the model's clean accuracy. Furthermore, our investigation into fine-pruning as a potential defense mechanism revealed that pruning alone is insufficient to remove the backdoor; effective mitigation requires extensive fine-tuning. These findings highlight a need for developing robust, routing-aware defenses tailored to sparse vision architectures.

An interesting direction for future work is extending our routing-aware attack approach to the text modality. While recent studies have successfully compromised MoE-based Large Language Models (LLMs) using text-based triggers, these attacks primarily operate at the token or phrase level. Consequently, they can suffer from the same fragmentation issues observed in vision models, where the textual trigger is not processed completely or is scattered across different experts. We hypothesize that text-based triggers will be more successful and require lower poisoning rates when they are explicitly optimized for the MoE architecture and its sequential expert routing dynamics.

\bibliographystyle{IEEEtran}
\bibliography{references}


\end{document}